\documentclass[11pt,a4]{article}
\usepackage{graphicx}
\usepackage{color}
\usepackage{ulem}

\title{\bf Extension of\\ the J-PARC Hadron Experimental Facility\\ - summary report -}

\date{April 30, 2017}
\author{}

\begin{document}

\maketitle

\begin{center}
\begin{Large}
Committee for the study of the extension of\\ the Hadron Experimental Facility
\end{Large}
\end{center}

\clearpage

The committee for the study of the extension of the Hadron
Experimental Facility was formed under the Hadron Hall Users'
Association in August, 2015.  
This document is a summary of the discussions among the committee members, 
and documented by a part of the
members listed below.

\vspace*{0.5cm}

\begin{tabular}{ll} 
Hiroyuki Fujioka &  (Kyoto U.) \\
Kenneth Hicks &    (Ohio U.)\\
Yoichi Igarashi   &  (IPNS, KEK)  \\
Kenta Itahashi   &  (Nishina Center, RIKEN)  \\
Takeshi~K. Komatsubara & (IPNS, KEK)  \\
Kouji Miwa & (Tohoku U.)  \\
Tomofumi Nagae & (Kyoto U.)  \\
Satoshi~N. Nakamura & (Tohoku U.)  \\
Hajime Nanjo$^{{*}}$ & (Kyoto U.)  \\
Hiroyuki Noumi & (RCNP, Osaka U. / IPNS, KEK)  \\
Hiroaki Ohnishi$^{{**}}$ & (Nishina Center, RIKEN / RCNP, Osaka U.) \\
Shinji Okada & (RIKEN)  \\
Atsushi Sakaguchi & (Osaka U.) \\
Shinya Sawada & (IPNS, KEK) \\
Suguru Shimizu & (Osaka U.) \\
Kotaro Shirotori & (RCNP, Osaka U.)  \\
Hitoshi Takahashi & (IPNS, KEK)  \\
Toshiyuki Takahashi & (IPNS, KEK)  \\
Hirokazu Tamura & (Tohoku U.) \\
Kiyoshi Tanida & (ASRC, JAEA)  \\
Mifuyu Ukai & (IPNS, KEK)  \\
\\
   & {*present address: Osaka U.} \\
   & {**present address: ELPH, Tohoku U.} \\
\end{tabular}

\clearpage 

This document is also supported by the following physicists.

\vspace*{0.5cm}

\begin{tabular}{ll}
Patrick Achenbach & (U. of Mainz)\\
Shuhei Ajimura    & (RCNP, Osaka U.)\\
Yuya Akazawa      & (Tohoku U.)\\
Claude Amsler     & (Stefan Meyer Institute for Subatomic Physics)\\
Kazuya Aoki  & (IPNS, KEK)\\
Hidemitsu Asano  & (RIKEN) \\
Sakiko Ashikaga  & (Kyoto U.)\\
Bernd Bassalleck  & (U.  New Mexico)\\
Elena Botta       & (Torino U. / INFN-Torino)\\
Wen-Chen Chang    & (Institute of Physics, Academia Sinica)\\
Hiroyuki Ekawa    & (Kyoto U.)\\
Alessandro Feliciello & (INFN-Torino)\\
Yu Fujii  & (Tohoku Medical and Pharmaceutical U.)\\
Manami Fujita     & (Tohoku U.)\\
David Gill        & (TRIUMF)\\
Toshi Gogami      & (Tohoku U.)\\
Yuji Goto         & (Nishina Center, RIKEN)\\
Philipp Gubler    & (Yonsei U.)\\
Taku Gunji        & (CNS, U. Tokyo)\\
Hiroyuki Harada  & (JAEA)\\
Toru Harada       & (Osaka EC U.)\\
Shoichi Hasegawa  & (JAEA)\\
Tadashi Hashimoto & (RIKEN)\\
Michael Hasinoff  & (U. British Columbia)\\
Shuhei Hayakawa   & (Osaka U.)\\
Satoru Hirenzaki  & (Nara Woman U.)\\
Ryotaro Honda     & (Tohoku U.)\\
Tomoaki Hotta     & (RCNP, Osaka U.)\\
Sanghoon Hwang    & (Korea Researh Institute of Standards and Science)\\
Masaya Ichikawa   & (Kyoto U.)\\
Yudai Ichikawa  & (JAEA)\\
Masaharu Ieiri    & (IPNS, KEK)\\
Masami Iio  & (ARL, KEK)\\
Kentaro Inoue     & (RCNP, Osaka U.)\\
Takatsugu Ishikawa & (ELPH, Tohoku U.)\\
Masahiko Iwasaki  & (RIKEN)\\
\end{tabular}

\clearpage 

\vspace*{0.5cm}

\begin{tabular}{ll}
Shunsuke Kanatsuki & (Kyoto U.)\\
Hiroki Kanda  & (Tohoku U.)\\
Masashi Kaneta  & (Tohoku U.)\\
Yohji Kato        & (IPNS, KEK)\\
Shinji Kinbara    & (Gifu U.)\\
Takeshi Koike     & (Tohoku U.)\\
Yusuke Komatsu    & (RCNP, Osaka U.)\\
Yoshihoro Konishi & (Tohoku U.)\\
Ami Koshikawa  & (Kyoto U.)\\ 
Geiyoub Lim       & (IPNS, KEK)\\
Toru Matsumura    & (National Defense Academy of Japan)\\
Michifumi Minakawa & (IPNS, KEK)\\
Yoshiyuki Miyachi & (Yamagata U.)\\
Yuhei Morino      & (IPNS, KEK)\\
Norihoto Muramatsu & (ELPH, Tohoku U.)\\
Ryotaro Muto$^{{*}}$  & (IPNS, KEK)\\
Sho Nagao         & (Tohoku U)\\
Fujio Naito  & (ACCL, KEK)\\
Yoshiyuki Nakada  & (Osaka U.)\\
Manami Nakagawa   & (Osaka U.)\\
Ken'ichi Nakano   & (Tokyo I. Technology)\\
Kazuma Nakazawa   & (Gifu U.)\\
Megumi Naruki  & (Kyoto U.)\\
Hidekatsu Nemura  & (U. Tsukuba)\\
Masayuki Niiyama  & (Kyoto U.)\\
Tadashi Nomura    & (IPNS, KEK)\\
Susumu Oda        & (Kyushu U.)\\
Yuji Ohashi       & (RCNP, Osaka U.)\\
Makoto Oka        & (Tokyo I. Technology / JAEA)\\
Haruhiko Outa     & (RIKEN)\\
Kyoichiro Ozawa   & (IPNS, KEK)\\
Jen-Chieh Peng    & (U. Illinois)\\
\end{tabular}

\clearpage

\vspace*{0.5cm}

\begin{tabular}{ll}
Takao Sakaguchi   & (BNL)\\
Yasuhiro Sakemi   & (CNS, U. Tokyo)\\
Hiroyuki Sako     & (JAEA)\\
Fumiaki Sakuma    & (RIKEN)\\
Masaharu Sato     & (RIKEN)\\
Susumu Sato       & (JAEA)\\
Yoshinori Sato    & (IPNS, KEK)\\
Michiko Sekimoto  & (RIKEN)\\
Toshiaki Shibata  & (Tokyo I . Technology)\\
Hajime Shimizu    & (ELPH, Tohoku U.)\\
Takao Shinkawa    & (National Defense Academy of  Japan)\\
Koji Shiomi       & (IPNS, KEK)\\
Myint Kyaw Soe    & (Gifu U.)\\
Yorihito Sugaya   & (RCNP, Osaka U.)\\
Hitoshi Sugimura  & (ACCL, KEK)\\
Yasuyuki Sugiyama & (ACCL, KEK)\\
Makoto Tabata     & (Chiba U.)\\
Yasuhisa Tajima  & (Yamagata U.)\\
Tomonori Takahashi & (RCNP, Osaka U.)\\
Minoru Takasaki   & (IPNS, KEK)\\
Makoto Takizawa   & (Showa Pharmaceutial U.)\\
Kazuhiro Tanaka  & (IPNS, KEK)\\
Manabu Togawa  & (Osaka U.)\\
Junji Tojo        & (Kyushu U.)\\
Atsuhi Tokiyasu   & (ELPH , Tohoku U.)\\
Shoko Tomita      & (Tohoku U.)\\
Yuichi Toyama     & (Tohoku U.)\\
Akihisa Toyoda    & (IPNS, KEK)\\
Hiroo Umetsu      & (Tohoku U.)\\
Hiroaki Watanabe  & (IPNS, KEK)\\
\end{tabular}

\clearpage

\vspace*{0.5cm}

\begin{tabular}{ll}
Takumi Yamaga     & (RCNP,  Osaka U.)\\
Hitoshi Yamamoto  & (Tohoku U.)\\
Takeshi Yamamoto  & (IPNS, KEK)\\
Taku Yamanaka  & (Osaka U.)\\
Shigehiro Yasui   & (Tokyo I. Technology)\\
Satoshi Yokkaichi  & (RIKEN)\\
Junya Yoshida     & (Gifu U.)\\
Tamaki Yoshioka   & (Kyushu U.)\\ 
\\
   & {*present address: ACCL, KEK} \\
\end{tabular}

\vspace*{0.5cm}

\clearpage

\tableofcontents

\clearpage

\section{Overview}

What are the most fundamental elements from which matter is made ? 
What laws of nature do the elements obey ? 
How is matter developed from them ? 
These are the questions that are at the heart of physics research. 
In particle and nuclear physics, 
we try to reveal the origin of matter, how the matter is developed, 
and the mechanism of the relevant phenomena. 
We summarize our main subjects as follows.

\begin{itemize}
\item Fundamental particles and interactions appeared in the early universe
at a time of high temperature
and are behind the physical laws of the universe. 
Through elucidation of these particles and interactions, 
we shed light on unanswered questions in the universe, 
such as the mechanism to realize the dominance of matter over anti-matter in the universe 
and the character of dark matter. 
\item Because the couplings among strongly-interacting quarks and gluons 
are complicated,
it is still unclear 
how hadrons, as a many-body system of quarks, 
and nuclear matter,  as a hadronic complex, 
are mathematically described. 
We investigate the nature of hadrons and nuclear matter 
in various environments, 
such as the high temperature in the early universe and the high density 
in the core of neutron stars, 
to clarify the mechanism which forms a variety of materials
such as the stars and human beings in the universe. 
\end{itemize}

\noindent
We are challenging these subjects 
with the high-intensity hadron beams produced
at the Hadron Experimental Facility of Japan Proton Accelerator Research Complex (J-PARC).

J-PARC provides the world highest-power beams for particle and nuclear experiments. 
J-PARC was constructed as the project with the highest priority of  the nuclear physics community in Japan
with strong support from the high energy physics community.
J-PARC attracts scientists from abroad 
and is an international core of excellence in Asia.

At present, 
two low-momentum charged-kaon beam lines (K1.8 and K1.8BR) 
and one neutral-kaon beam line are being operated 
in the north and south areas of the experimental hall, respectively, 
All of these beam lines are connected to the primary target that produces 
secondary particles. 
In the south area, a high-momentum beam line and a muon beam line will be 
ready for operation in the near future. 
From experiments with charged kaon beams,
fruitful results have been reported;
for example, 
candidates for deeply bound kaon-nuclear states, 
the discovery of charge symmetry breaking 
in the level structure of hypernuclei, 
and the observation of hypernuclei with two strange quarks.
With a neutral kaon beam, 
a search for the rare decay $K^0_L\rightarrow\pi^0\nu\bar{\nu}$, 
which is sensitive to new physics beyond the Standard Model of particle physics, is on the right track; 
this experiment has already achieved the world highest sensitivity to the branching fraction.

\vspace*{0.5cm}

A lot of results on nuclear and particle physics have been achieved
from the experiments at the Hadron Experimental Facility of J-PARC.
However,  we also recognized that an extension of the facility
is necessary for further researches in the future.

Only one production target is placed and is shared by the secondary beam lines
in the present hall. 
About 50 \% of the beam power of the primary proton from the J-PARC Main Ring accelerator
is used to produce secondary particles, 
and the remaining protons are transported to the beam dump. 
The secondary beam lines were built in the limited space of the present hall.
If the facility is extended to 
install the second and third production targets, and 
to construct new secondary beam lines for utilization of 
the high-power beam of J-PARC, 
we can expand our researches substantially.
The merits of the extension are as follows.

Firstly, 
the new beam lines will enable us to 
promote new measurements
which are impossible in the present facility.
For the strangeness nuclear physics and hadron physics
described in the following sections, 
construction of the high-intensity high-resolution beam line (HIHR) 
and the high-momentum mass-separated beam line (K10), 
with unprecedented capability, 
are highly anticipated. 
Since these new beam lines do not fit in the existing hall, 
the extension is necessary.

Secondly, 
we can improve the sensitivities of rare kaon-decay searches.
The extraction angle of the existing neutral beam line is chosen to be 16 degrees 
due to the limitation of the space of the present hall.
Extension of the hall enables us to
optimize the flux of kaons and neutrons in the neutral beam;
with an extraction angle of 5 degrees, 
we can utilize a more intense 
kaon beam and thereby dramatically increase the sensitivities.

Thirdly, 
more flexible operation of charged kaon beam lines will be realized. 
The existing K1.8 beam line is optimized for the studies of the nuclei with two strange quarks, 
and the studies for the nuclei with a single strange quark are
performed at the branch line (K1.8BR), which shares the upstream part
of the K1.8 beam line. 
Since those two beam lines cannot be operated at the same time, 
we are forced to perform the K1.8 and K1.8BR experiments one-by-one. 
With a new low-momentum beam line (K1.1), 
which provides kaon beams with higher quality (e.g., better $K/\pi$ ratio) and intensity, 
the experiments with two beam lines can be performed simultaneously.

Fourthly,
flexible operation can be realized also for high momentum beam lines.
In the extended hall, 
both the existing high-momentum beam line and a new high-momentum beam line (K10), 
which provides separated charged particles up to 10 GeV/$c$,
are placed in the south side, 
with the low momentum beam lines (K1.8, K1.1, and 
HIHR) in the north side. 
These five beam lines can be operated 
simultaneously and thus can accommodate the increasing demands
from users. 

\vspace*{0.5cm}

This report describes the project to extend the Hadron Experimental Facility and construct new beam lines.
These plans had already been considered in the initial plan of 
J-PARC conceptual designs,
and are updated and revised in this report to reflect recent progress.

This is a field-crossing project of particle and nuclear physics. 
The nuclear physics community in Japan supports it as the top-priority project,
and 
the particle physics community also recognized it as an important project.
The project was discussed in "Master Plan 2014" by 
the Science Council of Japan, and was 
reported in the revised version of "Roadmap 2014" issued in September 2015 by Ministry of Education, Culture, Sports, 
Science and Technology (MEXT) of Japan.
In this opportunity, we aim to motivate the project and then 
elucidate the mysteries in particle and nuclear physics 
described above.

\clearpage


\section{Facility}

At the Hadron Experimental Facility of J-PARC, 
the primary proton beam of 30~GeV is slowly extracted from the Main Ring accelerator
and transported to the production target (T1) in the experimental hall.
Various secondary particles such as K and $\pi$ mesons
produced in the target are transported through the secondary beam lines to the experimental area,
and are used for particle and nuclear physics experiments.
The construction of the facility was started in 2004,
and the first beam was extracted to the hall on January 27th, 2009.
The formal beam operation for users started from January 2010.

\begin{figure}[htbp] 
   \centering
   \includegraphics[width=9.0cm, angle=0]{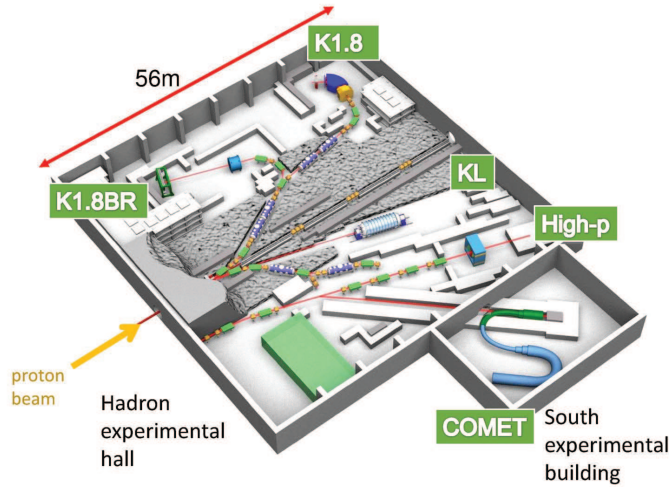} 
   \caption{Layout of the present experimental hall.}
   \label{fig:overview:layout_0}
\end{figure}

The layout of the present experimental hall is shown in
Fig~\ref{fig:overview:layout_0}.
The size of the hall is 58~m in width and 56~m in length,
and has a semi-underground structure with the height of 16~m above
and 6~m under the ground.
At present three secondary beam lines: 
K1.8 and K1.8BR beam lines in the north area
and KL beam line in the south area
are being operated.
Hadron-nuclear physics experiments are being carried out
by using charged $K^-$ mesons in the K1.8 and K1.8BR beam lines,
although these two beam lines cannot be operated simultaneously.
In the KL beam line, 
a particle-physics experiment using neutral $K_L$ mesons is being performed.
In the south area of the hall, in the next few years, 
a high-momentum (High-p) beam line as a branch of the primary proton beam
and a COMET beam line for a particle-physics experiment with muons in the South Experimental Building
will start the operation.

In the plan of facility extension,
the following new secondary beam lines will be constructed for new experiments
in order to expand the potential for particle and nuclear physics.

  \begin{enumerate}
     \item New beam lines optimized for physics goals
     \begin{description}
       \item [new KL beam line] :\ \  high-intensity neutral $K_L$ meson
       \item [HIHR beam line] :\ \  high-intensity and high-resolution charged $\pi$ meson
     \end{description}
     \item Beam lines to provide new secondary particles
     \begin{description}
       \item [K10 beam line] :\ \  high-momentum (2-10~GeV/$c$) charged $K$ meson, $\pi$ meson, and anti-proton
     \end{description}
     \item A beam line to extend $S=-1$ strange nuclear physics
      \begin{description}
       \item [K1.1 beam line] :\ \  low-momentum ($<$1.2~GeV/$c$) charged $K$ meson
     \end{description}
  \end{enumerate}

\begin{figure}[htbp] 
   \centering
   \includegraphics[width=13.0cm, angle=0]{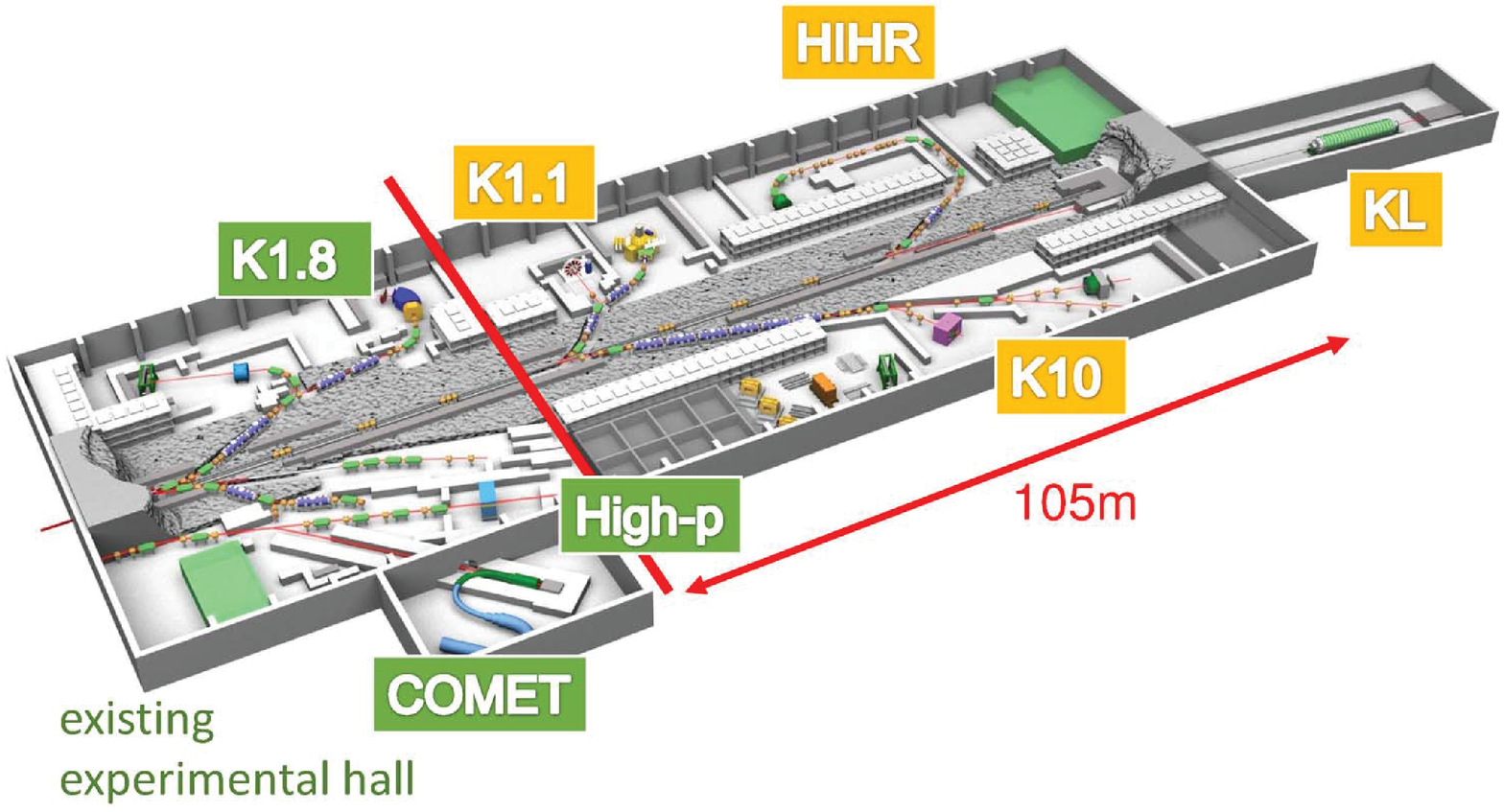} 
   \vspace*{0.2cm}
   \caption{Layout plan of the extended experimental hall}.
   \label{fig:overview:layout_1}
\end{figure}

The hall is extended to the downstream by 105~m
(Fig.~\ref{fig:overview:layout_1}).
The existing primary beam line (A line) is extended,
and two new production targets (T2 and T3) are installed.
The technologies necessary to extend the facility,
such as the beam optics to irradiate primary protons to two or more targets
and the movement of the existing beam dump to the downstream,
have  already been established.
The K1.1 and K10 beam lines are extracted from the T2 target station,
and the HIHR and new KL beam lines are extracted from the T3 target station.
The beam dump to absorb primary protons is located downstream of the T3 target.
The detector for the new $K_L$ experiment will be installed downstream of
the beam dump, so that the experimental area will be extended by 48~m.

The K1.8, High-p, and COMET beam lines that already exist in the hall
continue to be operated.
Thus, in the extended experimental hall, 
seven secondary beam lines in total can be operated.
We will push forward with the strangeness nuclear-physics experiments 
at the K1.8, K1.1, and HIHR beam lines in the north area,
the hadron-nuclear physics experiments at the high-p and K10 beam lines in the south area,
and particle-physics experiments at the COMET and new KL  beam lines.
Table~\ref{tab:beamlinesummary} summarizes the parameters
of these beam lines.
A test beam line is planned to be constructed in the experimental area
that is previously used for the KL experiment.

\begin{table}[htbp]
\caption{Beam lines in the Hadron Experimental Facility.}
\begin{center}
\begin{tabular}{|l||l|l|l|l|} \hline
          &  particle &  momentum  & number of particles & characteristics \\
 \hline
 \hline
 \multicolumn{5}{|l|}{beam lines in the present hall}\\
 \hline
         K1.8             &  $K^{\pm}$, $\pi^{\pm}$  & $<$ 2.0~GeV/$c$      & $10^{6}$ $K^{-} $/spill & {separated} \\
         K1.8BR       &  $K^{\pm}$, $\pi^{\pm}$  & $<$ 1.1~GeV/$c$      & {$10^{5}$ $K^{-} $/spill}  & {separated} \\
         KL                &  $K_L$                                & 2.1~GeV/$c$ in ave.  & {$10^{7}$ $K_L $/spill}  & {to 16 degrees}  \\
         High-p         &  p                                        &                                   &   {$10^{10}$  p /spill}         &  {primary protons}    \\
                              &  $\pi^{\pm}$                      &  $<$31~GeV/$c$       &   {$10^{7}$  $\pi$ /spill}  &        \\
  COMET              &  $\mu^-$                           &                                       &                                                  & {$\mu^- \to e^-$ conversion}\\
\hline
\hline
 \multicolumn{5}{|l|}{new beam lines in the extended area}\\
 \hline
         K1.1             &  $K^{\pm}$, $\pi^{\pm}$  & $<$ 1.2~GeV/$c$      & {$10^{6}$ $K^{-} $/spill}   & {separated}  \\
                              &                                             &  0.7$\sim$0.8~GeV/$c$ &                                               & {lower momentum} \\
                              &                                              &                                            &                                               &\ \ \ \ \ \ [\ K1.1BR\ ] \\
         HIHR           &  $\pi^{\pm}$                      & $<$  2.0~GeV/$c$      & $2.8\times 10^{8}$ $\pi$/spill & {separated} \\
                             &                                             &                                       &                                                  & {$\times$10} better $\Delta p/p$  \\
         K10              &  $K^{\pm}$, $\pi^{\pm}$, $\overline{p}$                & $<$  10~GeV/$c$      & {$10^{7}$ $K^{-} $/spill} &  {separated} \\

         new KL   &  $K_L$                              & 5.2~GeV/$c$ in ave.    & {$10^{8}$ $K_L $/spill}   & {to 5 degrees}  \\
                              &                                             &                                       &                                                  & n/$K_L$  optimized  \\
		\hline
\end{tabular}
\end{center}
\label{tab:beamlinesummary}
\end{table}%

The physics goals of the experiments in each beam line is described in the subsequent sections.
In brief, 
\begin{itemize}
\item strangeness nuclear physics  to understand many-body systems of hadrons.
 \begin{description}
       \item [K1.1 beam line] :\ \ to investigate the baryon-baryon interactions and the characteristics of baryons
              in the nuclear medium by using hyperons.
       \item [HIHR beam line] :\ \ to investigate the baryon-baryon interactions in high density
              by measuring the hyper-nuclear energy levels.
 \end{description}
\item hadron-nuclear physics to understand quark many-body systems.
 \begin{description}
       \item [K10 beam line] :\ \ to investigate the effective degrees of freedom
              to describe hadrons and the characteristics in the nuclear medium.
 \end{description}
\item particle physics to search for physics beyond the Standard Model.
 \begin{description}
       \item [new KL beam line] :\ \ to understand the dominance of matter over anti-matter 
              in the universe by detecting 100 events of CP violating rare decay.
 \end{description}
\end{itemize}

\clearpage


\section{Particle Physics}
The following two Nobel Prizes in Physics
illustrate the current issues on experimental particle physics related to the extension.
One is to Makoto Kobayashi and Toshihide Maskawa in 2008 
for their discovery, in 1973, 
of the origin of the broken CP symmetry which predicts the existence of at least three families of quarks in nature.
The other is to Fran\c{c}ois Englert and Peter W. Higgs in 2013
for the theoretical discovery of a mechanism that contributes to our understanding of the origin of mass of subatomic particles.

The Advanced Information in 2008~\footnote{
``The Nobel Prize in Physics 2008 - Advanced Information''.\\
{\it Nobelprize.org}. Nobel Media AB 2014. Web. 19 Apr 2016.\\
{\tt <http://www.nobelprize.org/nobel\_prizes/physics/laureates/2008/advanced.html>}
}
explains this award as: 
\begin{quotation}
 In 1967, Andrei Sakharov (the Nobel Prize in Peace in 1975) pointed out
 in a famous work that CP violation must be the cause of the asymmetry
 in the universe. It contains more matter than antimatter. The CP
 violation that the KM (Kobayashi-Maskawa) Model gives rise to is most probably not enough
 to explain this phenomenon. To find the origin of this CP violation we
 probably have to go beyond the Standard Model. Such an extension should
 exist for other reasons as well. It is believed that at higher energies
 other sectors of particles, so heavy that the present day accelerators
 have been unable to create them, will augment the model. It is natural
 that these particles will also cause CP violations and in the
 tumultuous universe just after the Big Bang these particles could have
 been created. These particles would have been part of the hot early
 universe and could have influenced it, by an as yet unknown mechanism,
 to be dominated by matter. Only future research will tell us if this
 picture is correct.
\end{quotation}

The Advanced Information in 2013~\footnote{
``The Nobel Prize in Physics 2013 - Advanced Information''.\\ 
{\it Nobelprize.org}. Nobel Media AB 2014. Web. 17 Apr 2016.\\
{\tt <http://www.nobelprize.org/nobel\_prizes/physics/laureates/2013/advanced.html>}
}
explains this award as: 
\begin{quotation}
 All measurements to date confirm that the properties 
 of the newly discovered particle are
 consistent with those expected for the 
 fundamental scalar boson predicted by the BEH (Brout-Englert-Higgs) mechanism.
 The discovery is a milestone for particle physics 
 and a tremendous success for the
 Standard Model. However, far from closing the book 
 it opens a number of new exciting
 possibilities: Theorists believe that the SM most probably 
 is but a low-energy approximation of a
 more complete theory. If this were not so, 
 quantum mechanical corrections to the Higgs mass
 would drive $m_H$ (the Higgs mass) towards the Planck scale 
 - unless “unnatural” cancellations occur. Therefore,
 extensions of the SM are proposed, keeping 
 the successful features of the SM but at the same
 time introducing “new physics” in a way, 
 which stabilizes $m_H$ at its low value, which is in
 accordance with SM expectations.
\end{quotation}

Both achievements are implemented in the framework of
the Standard Model (SM) of particle physics
and well describe the experiments of elementary particles.
On the other hand, 
there remains questions that cannot be explained by the SM, 
and the searches for new physics beyond the SM have been pursued intensively.
The questions include
the mechanism to produce the matter-dominant universe
and the origin of the low mass of the Higgs particle.
The CP-violation is 
a must for the matter-dominant universe,
but the size of the CP violation
implemented in the SM is too small to explain the
matter dominance in our universe.
A new mechanism should be introduced, and
new particles at a much higher energy-scale may contribute to the new breaking of CP symmetry.
Such particles would also provide the mechanism to stabilize the Higgs mass to the value observed.
In summary, 
the key to answer the questions is at the higher energy-scale.

To search for new physics beyond the SM, 
there are two approaches in the accelerator-based particle physics experiments.
One is the Energy Frontier,
such as the ATLAS and CMS experiment in the LHC,  
which aims to produce new particles directly
by colliding the beams to maximize the center-of-mass energy.
The other is the Intensity Frontier, 
in which 
a high energy scale is realized in a short time-period
due to the Uncertainty Principle in Quantum Mechanics between time and energy.
Heavy new particles can contribute to rare processes 
and make phenomena different from the SM predictions.
High intensity beams are necessary
in the experiments to study
rare processes to see a golden needle in a haystack
and to explore new physics beyond the SM.

In the search for new physics,
the usual processes caused by the SM are considered to be the background.
Thereby, if such SM processes were suppressed 
due to symmetry principles,
new physics effects can be enhanced because they do not need to follow
the suppression mechanisms of the SM.
Furthermore,
the processes should be predicted by the SM as precisely as possible; 
when we observe a deviation from the SM in a process,
it is not so meaningful if the deviation is within the uncertainty of the SM prediction.

The particle-physics experiments performed in the Hadron Experimental Facility of J-PARC
are exactly what we pursue in the Intensity Frontier.
In the present hall,
two experiments using 
K mesons are being conducted;
they are directly connected to the extension of the facility.
Also, a new COMET experiment to search for $\mu$-e conversion is being prepared.
They are complementary to the experiments using B mesons in other facilities; 
by comparing their physics effects
the new physics scenario can be identified.

One of the kaon experiments at the present hall is
the E14 KOTO experiment
to search for the decay $K_L\rightarrow \pi^0\nu\bar{\nu}$,
whose branching-fraction prediction in the SM is very small,
$3\times 10^{-11}$, and thus
the decay has never been observed.
Since this decay is 
strongly suppressed, due to the hierarchy at the flavor transitions
in the Kobayashi-Maskawa matrix,
it is sensitive to the new physics processes that does not obey such suppression.
This decay is also a CP-violating process and
a new source of CP violation can be explored.
The uncertainty in the SM prediction, 2\%, is small,
and some new physics theories predict a branching fraction of
an order-of-magnitude larger.
The KOTO experiment will search at the sensitivity of
$\mathcal{O}(10^{-11})$ 
in order to detect effects from new physics.
In the extended facility, 
a new beam line and a new detector will be constructed
aiming to measure the branching fraction of the decay
with higher statistics
as the KOTO step-2 experiment. 
The step-2 experiment plans
to detect $\mathcal{O}(100)$ events of this decay
even with a branching ratio similar to that in the SM.

The other kaon experiment at the present hall is is 
the E36 experiment which uses charged-kaon decay to explore
the breaking of
of the universality of lepton flavor.
In the experiment,
the ratio of the branching fractions of
the $K^+\rightarrow \mu\nu_{\mu}$ and $K^+\rightarrow e\nu_{e}$ decays 
is measured and
is compared to the SM expectation.
The decay $K^+\rightarrow e\nu_{e}$ is suppressed due to the helicity suppression 
in the structure of the weak interaction in the SM.
The sensitivity to new physics is enhanced
with the ratio of the two branching fractions,
since 
the theoretical as well as systematic uncertainties are reduced. 
In the extended facility,
the collaboration plans to perform
the E06 TREK experiment
to study the breaking of time-reversal (T) symmetry
with the same spectrometer magnet.
The transverse polarization of the muon
from the $K^+\rightarrow \pi^0\mu^+\nu$ decay in the decay plane
is measured.
Since 
the polarization is a T violating observable,
a polarization larger than the SM prediction 
would indicate T violation.
T violation is connected to the CP violation through the CPT theorem,
and is expected to be a key to solve the mystery of a matter-dominant universe.

\clearpage
\section{Hadron and Nuclear Physics}

“What is the origin of all the matter in the universe?” - this is a simple question 
which humankind has asked since the ancient times.  
The research with high energy accelerators in the 20th century
found that the six leptons
 (an electron and its sisters, and neutrinos) 
 and the six quarks are the elementary particles which cannot be divided anymore.  
 The interactions between these particles are caused 
 by exchanging gauge bosons.
 The strong interaction between quarks is carried by gluons and described by the quantum chromo-dynamics (QCD).  
 Hadrons, such as protons and neutrons, are made from quarks, and atomic nuclei are made of protons and neutrons.  
 Atomic nuclei make up atoms and molecules by capturing electrons, 
 and then 
 various kinds of materials, which are constituent of stars and humankind, are produced.  
 This hierarchy traces the history of the creation of materials of the universe 
 from its beginning, starting with the Big Bang.  
 One of the most important issues of modern hadron and nuclear physics is 
 to investigate various aspects of particles with quark multi-body systems 
 and to understand mechanisms of their production.

Because quarks are confined in hadrons, 
there is not enough knowledge on how hadrons are made up from quarks and on the characteristics 
of the quark/hadronic systems, 
depending on the temperature and density.  
Thus, the project of extending the Hadron Experimental Facility addresses three issues; 
the first is 
“how hadrons are produced and excited”, 
the second is 
“understanding of the baryon-baryon interaction which makes atomic nuclei”, 
and the third is 
“understanding of baryonic matter with super-high density - why heavy neutrons stars exist”.  
In particular, on the second and third issues, 
understanding of the short-range interaction is a key 
to the characteristics of the high-density hadronic matter, and
is the major subject.

In the following subsections, 
the overview and plans of addressing these issues are described.

\subsection{how hadrons are produced and excited}

The dynamics of quarks and gluons are described by the theory of quantum chromodynamics (QCD). 
The coupling constant of QCD strongly depends on its energy, 
which introduces  a scale parameter, i.e. $\Lambda_{QCD}$. 
The strength of the parameter is known to be about 0.2~GeV. 
The physical picture of the strongly interacting matter is very different below and above $\Lambda_{QCD}$. 
Above $\Lambda_{QCD}$, phenomena can be described well by perturbative QCD.
However, below $\Lambda_{QCD}$, due to the large coupling constant, non-perturbative effects cannot be neglected. 
Two characteristic phenomena appear in this low energy QCD regime. 
One is “confinement” and the other is “generation of the dynamical mass of the hadron”. 
One of the ways to solve low energy QCD is indeed the lattice QCD calculations. 
Nowadays, the lattice QCD calculations are well developed and 
mass spectra of ground state hadrons are reproduced. 
However,  mass spectra for excited state hadrons still cannot be described. 
On the other hand, a model based on constituent quarks (a.k.a. the quark model) 
also succeeds in describing ground state hadrons;
however, excited-state baryons, such as $\Lambda$(1405) and Roper resonance(N*(1440)), cannot be described very well. 
Recently, new exotic hadrons, i.e. $X$, $Y$, $Z$ and penta-quark hadrons ($\Theta^+$ and $P^+_c$), are discovered in highly excited hadrons. 
Those states also cannot be described with the lattice QCD calculations, 
nor with the simple quark model. 
Thus none of the models or theories can reach a perfect description of the hadron spectra including excited-state hadrons. 
To understand hadron spectra including the excited-state hadrons, 
we need to shed light on the effective degrees of freedom inside the hadrons and the mechanism of 
how the hadrons are excited.

One of the ways to understand the dynamics inside a baryon is 
to replace one of two light-quarks in the baryon to heavy quarks, 
such as a strange quark and/or a charm quark. 
The color magnetic interaction is proportional to the $1/m_q$, 
and the interaction between a heavy-quark and a light-quark is going to be reduced. 
Thus the correlation between light quarks (di-quark correlation) in a baryon will be enhanced. 
The strength of the correlation would depend not only on the baryon-excited energy 
but also on the decay branch and the production cross section. 
At J-PARC, spectroscopic studies of charmed baryon and multi-strangeness baryon ($\Xi^*$ and $\Omega^*$) will be performed.

Spontaneous breaking of chiral symmetry is one of the fundamental concepts of hadron physics
and characterizes the QCD vacuum structure. 
The restoration of chiral symmetry is predicted in high-density or high-temperature QCD matter. 
It is easily expected that the properties of hadrons, such as the mass and decay width, 
will be modified under such extreme conditions. 
However, the pattern of the modification is predicted to be different in different hadrons 
(in particular for mesons). 
At J-PARC, precise measurements of meson spectra in the nuclear matter will be performed 
with light vector mesons ($\phi$, $\rho$, $\omega$), 
pseud-scalar mesons ($\eta$, $\eta^\prime$) and light-heavy mesons ($D$ mesons). 
Comparisons of the in-medium modification of these mesons will reveal the mechanism of partial restoration of chiral symmetry in nuclear matter. 
Precise measurements of the light vector-meson spectral functions in nuclei will be performed in the High-p beam line. 
Extending the facility will open a new horizon, 
i.e. $D$ mesons in nuclei by the measurements 
of $D-\overline{D}$ meson production near the production threshold 
using an intense and high-energy anti-proton beam. 
Understanding of the meson property in nuclei and/or nuclear matter is needed
to describe hadrons in the extremely high-density matter, i.e. the core of neutron stars.

\subsection{understanding of the baryon-baryon interaction which makes atomic nuclei}

Nucleon scattering experiments have indicated that the nuclear force 
is strongly repulsive at ranges shorter than 1~fm. 
Due to small nucleon momenta, 
the structure of ordinary nuclei is determined mainly by the strength of the nuclear force  
averaged over short and long ranges and is not affected by details of the short-range parts. 
Theoretical models for the nuclear force are 
based on the meson exchange picture proposed by Yukawa about eighty years ago. 
They include a phenomenological short-range repulsive core at ranges shorter than 1~fm 
where the two nucleons overlap with each other. 
So far, these models have succeeded when they are used to describe nuclei.

In contrast, the short-range parts of the nuclear force play essential roles 
in the high-density nuclear matter realized in the core of neutron stars. 
It would be natural to describe the nuclear force using the degrees of freedom 
of quarks and gluons rather than the meson exchange picture at the ranges shorter than 1~fm, 
where two nucleons are overlapping and their structure affects the nuclear force. 
In addition, since the Fermi energy of nucleons in high-density nuclear matter increases 
with density, 
hyperons are expected to appear at a certain density. 
Thus, the nuclear force should be generalized to other baryon-baryon interactions 
(all the forces between the baryon octet members including hyperons) 
in studies of high-density nuclear matter.

Theoretical studies on the baryon-baryon interactions based on quark models suggest that 
the strength of repulsive cores depends on the spin and flavor of the two baryons. 
It is expected that there is a very large repulsive core in a specific state, such as $\Sigma^+p$ 
in the spin singlet channel, 
due to the Pauli effect between quarks, 
while there would be an attractive core due to the color-magnetic interaction by one-gluon exchange in the $H$ dibaryon channel. 
The spin-orbit interaction, which is important at short ranges, has been, so far, 
explained with heavy meson exchange.
However, if the spin-orbit force is due to gluon exchange, 
a quite-different behavior from the nucleon-nucleon case is expected 
in the spin-orbit force of the hyperon-nucleon ($YN$) interaction. 
These quark-model predictions are in qualitative agreements with the baryon-baryon interactions obtained from lattice QCD simulations.

Existing hyperon-nucleon scattering data have been limited 
in the measured momentum ranges as well as in the amount of statistics.
Nevertheless, 
the baryon-baryon interaction models based on a one-meson exchange picture 
have been developed by using spectroscopic information from 
hyper-nuclear structure and 
by supplementing the scarce hyperon-nucleon data 
with more plentiful nucleon-nucleon scattering data and with the help of flavor $SU$(3) symmetry. 
These models, unfortunately, have no predictive power for the short-range parts of the hyperon-nucleon interaction, 
although they can reproduce the $YN$ scattering data and hyper-nuclear information rather well. 
Since the existing $YN$ scattering data are not so sensitive to the short-range parts of interaction, 
new data are essential to reveal these interesting features.

In this situation, 
we plan to introduce the most advanced techniques to the experiments and drastically improve the $YN$ scattering data 
both in quality and in quantity. 
The new scattering data, also compared with lattice QCD calculations, 
enables us to establish the baryon-baryon interaction models covering the short-range parts. 
As described in the next subsection, the $YNN$ three-body repulsive force 
that cannot be reduced into the two-body baryon-baryon interactions 
is considered to play an important role in describing the core of neutron stars. 
Experimental verification of the three-body force effects also requires 
well-established two-body baryon-baryon interaction models covering the short-range parts.

\subsection{understanding of baryonic matter with super-high density - why heavy neutrons stars exist}

Nucleus can be characterized with features such as constant density (saturation of density) 
and constant binding-energy per nucleon (saturation of binding energy). 
It is an important scientific issue to deduce these characteristics of nuclei 
from the nuclear force models, and 
many efforts have been paid to solve it. 
It has been clarified that the two-body nuclear force is not enough and
the introduction of three-body forces is essential 
to reproduce high-precision experimental results of nuclear binding energies.
Recent theoretical calculations indicate that the three-body nuclear force is repulsive in heavy nuclei 
and it should give more significant repulsion in the higher-density nuclear matter 
such as deep inside of neutron stars. 
Modern theoretical calculations predict that 
the three-body repulsive baryonic-force plays an important role in deciding the maximum allowed mass of neutron stars.

The Fermi energy of neutrons becomes quite high deep inside of neutron stars, 
so that a neutron converts to a hyperon such as a $\Lambda$ particle to reduce the total energy 
when the density of a neutron star becomes higher than a threshold value. 
The interaction between a $\Lambda$ and a nucleon is attractive as we learned from various studies of hypernuclei. 
Thus, neutrons convert to $\Lambda$ particles 
when the density becomes larger than a few times of the normal nuclear density. 
We naturally expect that hyperons exist deep inside of neutron stars. 
Rich NN scattering data enable us to construct realistic nuclear force models. 
Spectroscopy of hypernuclei gives information on the $YN$/$YY$ interactions. 
The internal structure of a neutron star is theoretically studied 
with the equation of state for high-density nuclear (hadronic) matter 
using realistic nuclear models with the knowledge of the $YN$/$YY$ interactions. 
Established models predict that hyperons such as $\Lambda$ particles appear in the central part of neutron stars and 
a predicted heaviest neutron star has about 1.5 solar mass. 
It had been consistent with the results of neutron mass measurements for long time, 
however two neutron stars with two solar mass were observed after 2010. 
Based on conventional theoretical models, such a heavy neutron star should collapse to a black hole 
because the equation of state of the star becomes too soft if it contains hyperons. 
The discrepancy between the observation and theoretical prediction cannot be solved 
even if the three-body nuclear force is taken into account. 
This is called as a “hyperon puzzle” and is a big challenge to nuclear physics researchers. 
It reveals that something important is missing in our understanding of hadronic interactions. 
The equation of state for high-density hadronic matter should be revised and updated.

A promising approach to solve the hyperon puzzle is 
to introduce a three-body repulsive force in the baryonic interaction with hyperons. 
This is a natural extension of the three-body repulsive nuclear force 
which is known to be necessary in the experiments. 
We intend to prove the existence of this three-body repulsive baryonic force by high precision experiments at J-PARC. 
Even if the experiments clarify that such three-body force is not the source of the hyperon puzzle, 
it would open a door to more exotic scenario such as the co-existence of nuclear matter and quark matter.

A three-body repulsive baryonic force affects the single particle energy of a hyperon 
which is bound deep inside of a hypernucleus. 
An effect on the $\Lambda$ binding energy is expected to be about a half MeV and 
it is smaller than the achievable accuracy of the established hypernuclear spectroscopic techniques with meson beams. 
The HIHR beam line, which will be newly constructed in the extended hall,
enables us to measure hypernuclei with ten times better accuracy ($<$0.1 MeV) by using the dispersion matching technique.
High precision spectroscopy can be expanded from light hypernuclei (A$\sim$10) to heavy ones (A $>$ 200). 
Such a series of systematic measurements will reveal the existence of a three-body repulsive baryonic force 
and solve the hyperon puzzle of heavy neutron stars.

\clearpage

\end{document}